\documentclass[aps,prl,twocolumn,showpacs,superscriptaddress,groupedaddress]{revtex4}  % for review and submission
\usepackage{graphicx}  % needed for figures
\usepackage{dcolumn}   % needed for some tables
\usepackage{bm}        % for math
\usepackage{amssymb}   % for math

\begin{document}
% The following information is for internal review, please remove them for submission

% the following line is for submission, including submission to the arXiv!!
\hspace{5.2in} %\mbox{Fermilab-Pub-04/xxx-E}

\title{Stochastic Formation of Polariton Condensates in Two Degenerate Orbital States}
% remove these 3 lines before journal submittal.
%\centerline{author list dated 8 August 2012}
% end removal before journal submittal
%
\affiliation{National Institute of Informatics, 2-1-2 Hitotsubashi, Chiyoda-ku, Tokyo 101-8430, Japan}
\affiliation{NTT Basic Research Laboratories, 3-1 Morinosato-Wakamiya Atsugi-shi, Kanagawa 243-0198, Japan}
\affiliation{Graduate School of Information Science and Technology, The University of Tokyo, 7-3-1 Hongo, Bunkyo-ku, Tokyo 113-0033, Japan}
\affiliation{Edward L. Ginzton Laboratory, Stanford University, Stanford, California 94305-4085, USA}
\affiliation{Technische Physik and Wilhelm-Conrad-Rontgen-Research Center for Complex Material Systems, Universitat Wurzburg, D-97074 Wurzburg, Am Hubland, Germany}
\author{Kenichiro Kusudo} \affiliation{National Institute of Informatics, 2-1-2 Hitotsubashi, Chiyoda-ku, Tokyo 101-8430, Japan} \affiliation{NTT Basic Research Laboratories, 3-1 Morinosato-Wakamiya Atsugi-shi, Kanagawa 243-0198, Japan} \affiliation{Graduate School of Information Science and Technology, The University of Tokyo, 7-3-1 Hongo, Bunkyo-ku, Tokyo 113-0033, Japan}
\author{Na Young Kim} \affiliation{Edward L. Ginzton Laboratory, Stanford University, Stanford, California 94305-4085, USA}
\author{Andreas L$\ddot{\rm o}$ffler} \affiliation{Technische Physik and Wilhelm-Conrad-Rontgen-Research Center for Complex Material Systems, Universitat Wurzburg, D-97074 Wurzburg, Am Hubland, Germany}
\author{Sven H$\ddot{\rm o}$fling} \affiliation{Technische Physik and Wilhelm-Conrad-Rontgen-Research Center for Complex Material Systems, Universitat Wurzburg, D-97074 Wurzburg, Am Hubland, Germany}
\author{Alfred Forchel} \affiliation{Technische Physik and Wilhelm-Conrad-Rontgen-Research Center for Complex Material Systems, Universitat Wurzburg, D-97074 Wurzburg, Am Hubland, Germany}
\author{Yoshihisa Yamamoto} \affiliation{National Institute of Informatics, 2-1-2 Hitotsubashi, Chiyoda-ku, Tokyo 101-8430, Japan} \affiliation{Edward L. Ginzton Laboratory, Stanford University, Stanford, California 94305-4085, USA}
 
%
% visitor_addresses.tex                       8 August 2012 
%  available symbols are:
%  $\ast, \dag, \ddag, \S, \P, $\|$, $\ast\ast$, \dag\dag, \ddag\ddag ,\#
%
%\collaboration{The D0 Collaboration\footnote{with visitors from
%{alton}
%$^{a}$Augustana College, Sioux Falls, SD, USA,
%{burdin}
%$^{b}$The University of Liverpool, Liverpool, UK,
%{garcia-guerra}
%$^{c}$UPIITA-IPN, Mexico City, Mexico,
%{grohsjean}
%$^{d}$DESY, Hamburg, Germany,
%{partridge}
%$^{e}$SLAC, Menlo Park, CA, USA,
%{hesketh}
%$^{f}$University College London, London, UK,
%{luna-garcia}
%$^{g}$Centro de Investigacion en Computacion - IPN, Mexico City, Mexico,
%{podesta-lerma}
%$^{h}$ECFM, Universidad Autonoma de Sinaloa, Culiac\'an, Mexico
%and
%{santos}
%$^{i}$Universidade Estadual Paulista, S\~ao Paulo, Brazil.
%{falkowski}
%$^{?}$Laboratoire de Physique Theorique, Orsay, FR,
%{hooper}
%$^{?}$Visitor from Bradley University, Peoria, IL, USA.
%{kozminski}
%$^{?}$}Visitor from Lewis University, Romeoville, IL, USA.
%{weber}
%$^{?}$Universit{\"a}t Bern, Bern, Switzerland.
%{deceased}
%$^{\ddag}$Deceased.
%}} \noaffiliation
%
\vskip 0.25cm
       % D0 authors (remove the first 3 lines
                             % of this file prior to submission, they
                             % contain a time stamp for the authorlist)
                             % (includes institutions and visitors)
\date{\today}

\begin{abstract}
We explore the exciton-polariton condensation in the two degenerate orbital states.
In the honeycomb lattice potential, at the third band we have two degenerate vortex-antivortex lattice states at the inequivalent K and K$'$-points.
We have observed energetically degenerate condensates within the linewidth $\sim $ 0.3 meV, and directly measured the vortex-antivortex lattice phase order of the order parameter.
We have also observed the intensity anticorrelation between polariton condensates at the K- and K$'$-points.
We relate this intensity anticorrelation to the dynamical feature of polariton condensates induced by the stochastic relaxation from the common particle reservoir. 
\end{abstract}

\maketitle
%\section{\label{sec:level1}First-level heading}
% sections are not used for PRL papers
Microcavity exciton-polaritons are a solid state system where bosonic condensation \cite{Pitaevskii} has been explored.
A semiconductor microcavity with embedded quantum wells (QWs) realizes the strong coupling between QW excitons and microcavity photons, which results in the new eigenmodes of the system called upper and lower polaritons \cite{Kavokin,Snoke}.
At low densities, polaritons are regarded as bosonic particles, and lower polaritons (LPs) are expected to exhibit dynamic condensation \cite{Imamoglu}.  
LP condensation has been observed in inorganic \cite{Deng,Kasprzak,Balili,Christopoulos} and organic \cite{Kena-Cohen} semiconductors.
One unique feature of polariton condensates comes from their non-equilibrium nature.
Because polaritons leak from the cavity as emitted photons before reaching the thermal equilibrium, the stochastic relaxation plays a role for the formation of the condensate.
Recently the stochastic polarization build-up of the polariton condensation has been theoretically and experimentally explored \cite{Read,Ohadi}.

Using the thin metal deposition method, single trap \cite{Utsunomiya}, one-dimensional array \cite{Lai} and 2D square lattice potentials \cite{Kim} have been studied, where the dynamical and bottleneck condensation in meta-stable higher bands has been observed.
A band gap between the bands suppresses a LP relaxation to the lower band, which results in the meta-stable condensation at the top of the gap.
Due to the short lifetime of the LPs, they can escape from the cavity as emitted photons prior to the relaxation to the lower bands, which allows the direct access to the particle density and phase distribution of the higher band condensates.
The mode competition and the relaxation dynamics for different energy modes have been studied experimentally and simulated by the rate equations \cite{Lai, Kim}.
For the degenerate energy states, the relaxation to these states are initiated by the independent spontaneous scattering and the stochastic population build-up will be induced between these states.
Here we report the stochastic formation of polariton condensates in two degenerate orbital states in the honeycomb lattice potentials.

\begin{figure}
\begin{center}
\includegraphics[scale=0.4]{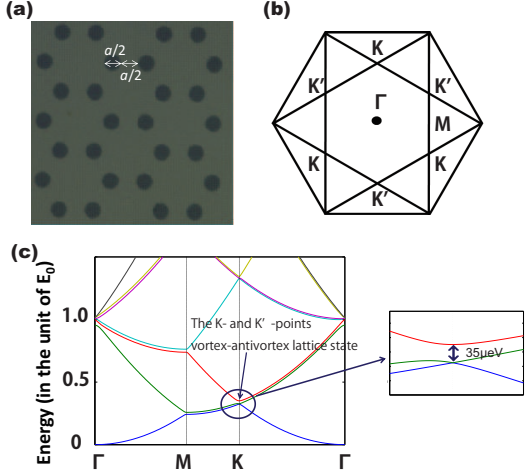}
\caption{(a) A photograph of the sample surface. Brighter areas where the thin metal films are
deposited correspond to the high potential regions, while darker areas with no metal
correspond to the polariton traps with low potential.
(b) Corresponding Brillouin Zones (up to the 3rd) and high symmetry points. (c)
Calculated band structures with the potential amplitude $V=0.15E_0$, where
$E_0=\hbar^2(2\pi)^2/2ma^2$.}
\label{Fig1}
\end{center}
\end{figure}

The honeycomb lattice potential is implemented by the thin metal deposition method with $\sim$ 150 $\mu$eV potential strength and the period $a \sim$ 2 $\mu$m (Fig. \ref{Fig1}(a)).
The characteristic kinetic energy is $\frac{\hbar ^2{k_0}^2}{2m_p} \sim 1.5$ meV with the polariton mass $m_p$, the reduced Plank constant $\hbar$ and the unit wavenumber $k_0=\frac{2}{\sqrt{3}}\frac{2\pi}{a}$ at a red detuned area ($\Delta \sim$ -2 meV).
The Brillouin Zones (BZs) with three high symmetry points ($\Gamma$, M and K) are shown in Fig. \ref{Fig1}(b). The band structure (Fig. \ref{Fig1}(c)) is calculated with single-particle plane wave bases for the weak potential regime.
There are two inequivalent K- and K$^{\prime}$- points at the vertices of the first BZ with 3-fold rotational symmetry.
Under the weak periodic potential, this 3-fold degeneracy is split into lower degenerate doublets and upper non-degenerate singlet with an energy gap ($\sim$ 35 $\mu$eV with our potential), which suppresses the LP relaxation from the upper singlet to the lower doublet states.

The details of our sample with 12 GaAs QWs in AlGaAs/AlAs planer microcavity are shown in the previous paper \cite{Kim}.
All measurements in this paper are performed at $\sim$ 4 K, and around a red-detuned area with the detuning $\Delta \sim$ -2 meV.
LPs are created by Ti:Sapphier laser of $\sim$ 767.5 nm (QW exciton resonance) incident at $\sim$ 60 degree in the pulsed mode with a 76 MHz repetition rate and $\sim$ 3 ps pulse width.

\begin{figure}
\begin{center}
\includegraphics[scale=0.4]{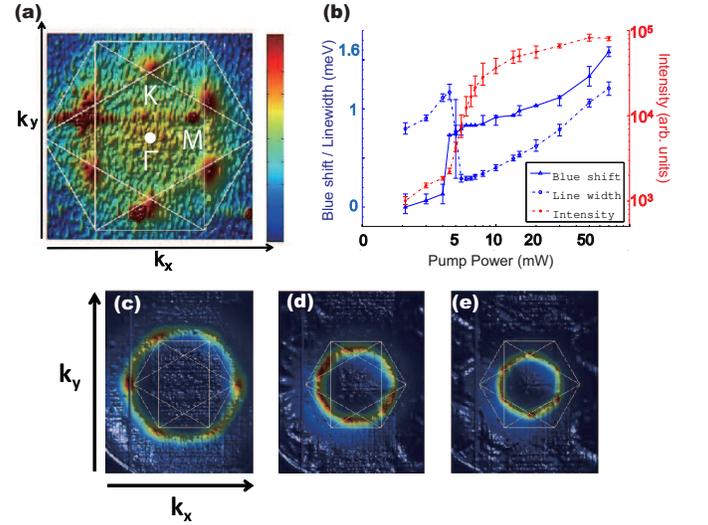}
\caption{(a) A time-integrated far-field image (in linear scale) showing LP condensation in momentum space
above the threshold pump power; P/P$_{\rm th}$ $\sim$ 2. The white lines indicate the BZs, which are
calculated with the diffraction signals of the reflected laser. (b) The pump power
dependence of the LP population, the energy shift at k=0 measured from the value below the threshold P = 2.5 mW and the linewidth of the K-point condensate. All the values are averaged over the six K and K$^{\prime}$-points, and the errorbars show the minimum and the maximum value of them. (c) (d) (e) Energy-resolved time-integrated far-field images at 1612.06 meV (c), at 1611.88 meV (d) and at the K-point condensate energy 1610.49 meV (e) taken at P/P$_{\rm th}$ $\sim$ 2.}
\label{Fig2}
\end{center}
\end{figure}

We first explore the far-field polariton distribution.
Sharp intensity peaks in the far-field pattern shown in Fig. \ref{Fig2}(a) indicate that large LP population condenses above the threshold pump power P/P$_{\rm th}$ $\sim$ 2 at the K- and K$^{\prime}$-points in the 3rd BZ.
We note that LPs are more or less equally populated at all six points within $\sim$15$\%$ difference.
At the K-points, the lowest three orthonormal eigenstates are split into two degenerate doublet $|\Psi^K_1\rangle$ and $|\Psi^K_2\rangle$ and the upper singlet $|\Psi^K_3\rangle$ \cite{Note1}.
The pump power dependence of the K-point energy, linewidth and intensity are shown in Fig. \ref{Fig2}(b).
Across the threshold pump power (P$_{\rm th}$ $\sim$ 4 mW), the intensity increases nonlinearly. The energy is blue-shifted at the threshold, and gradually increases with increasing the pump power.
The linewidth shrinks at the threshold and increases above the threshold due to the polariton-polariton interaction.
All of above results show the characteristic behavior of polariton condensation \cite{Deng,Kasprzak,Sup1}.
It is also experimentally confirmed that the energy of all six K- and K$^{\prime}$-points are degenerate within their spectral linewidth ($\sim$ 0.3 meV).
The energy-resolved far-field patterns reveal the relaxation path of LPs (Figs. \ref{Fig2}(c)-(e)).
From the 2nd $\Gamma$-point to the K(K$'$)-points, LPs relax through the 3rd BZ not the 2nd BZ, which is consistent with the condensation at the 3rd band K(K$'$)-points.
Because LPs relax by polariton-polariton interaction and polariton-acoustic phonon interaction in low temperatures both of which conserve the in-plane momentum, it is natural to think that LPs relax to either $|\Psi^K_3\rangle$ or $|\Psi^{K'}_3\rangle$ state (not superposition states of two).
\begin{figure*}
\begin{center}
\includegraphics[scale=0.3]{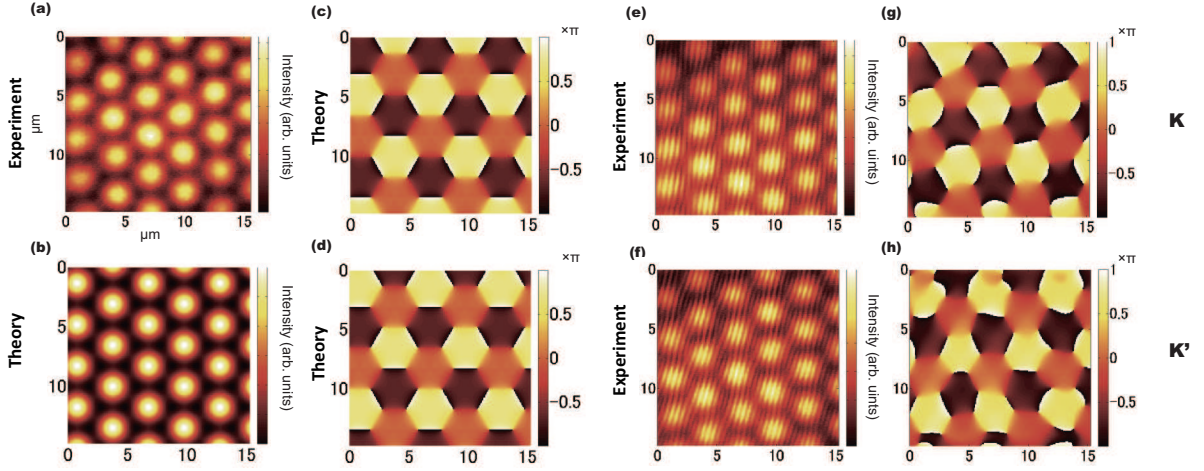}% Here is how to import EPS art
\caption{
The time-integrated interferogram for the K- and K$^{\prime}$-points condensate at P/P${\rm _{th}} \sim$ 2.
(a) The near-field image which includes all the K and K$^{\prime}$-points signal. (b) The theoretical near-field image of $|\Psi^{K}_3\rangle$ and $|\Psi^{K'}_3\rangle$.
(c) (d) Theoretical phase profile of $|\Psi^{K}_3\rangle$ (c) and $|\Psi^{K'}_3\rangle$ (d).
(e) (g) The experimental interferogram between the signal from all the K and K$^{\prime}$-points in the signal arm, and the signal from only one K-point in the reference arm (e) and the extracted phase profile of its off-axis component (g).
(f) (h) The same interferograms, but for the K$'$-points.}
\label{Fig3}
\end{center}
\end{figure*}

The order parameter of $|\Psi^K_3\rangle$ and $|\Psi^{K'}_3\rangle$ condensates are directly measured by a modified Mach-Zehnder interferometer \cite{Sup2}.
In a signal arm, signals from all six K- and K$^{\prime}$-points are chosen.
In a reference arm, we choose one of the three K- or K$^{\prime}$-points using a 150 ${\rm \mu}$m pinhole (corresponding to $\Delta k \sim 0.31$ ${\rm \mu m^{-1}}$) as a reference plane wave for the interferogram.
We use a band-pass filter with 1 nm pass band width to filter out the residual light from other states as well as the scattered laser.
The near-field image observed in the signal arm shown in Fig. \ref{Fig3}(a) is consistent with the theoretical particle density distribution of $|\Psi^{K}_3\rangle$ and $|\Psi^{K'}_3\rangle$ in Fig. \ref{Fig3}(b) with the intensity peak-to-peak distance of $2\sqrt{3}$ ${\rm \mu m}$.
We note that $|\Psi^{K}_3\rangle$ and $|\Psi^{K'}_3\rangle$ have the same intensity distribution.
This result shows that $|\Psi^{K}_3\rangle$ and $|\Psi^{K'}_3\rangle$ are incoherent in the time-averaged sense, because superposition states have the different intensity distribution \cite{Sup3}.
The theoretical phase profiles of $|\Psi^{K}_3\rangle$ and $|\Psi^{K'}_3\rangle$ are shown in Fig. \ref{Fig3}(c) and Fig. \ref{Fig3}(d) respectively, where clockwise and counter-clockwise phase rotation exist to form the honeycomb lattice geometry.
At the position where the phase rotation exists the intensity goes to zero (in Fig. \ref{Fig3}(b)).
This vortex-antivortex lattice order is directly detected in the experimental interferogram shown in Figs. \ref{Fig3}(e) and (f).
By extracting the off-axis components of the interferogram \cite{Lagoudakis1,Nardin2}, we can reconstruct the phase profiles of the $|\Psi^{K}_3\rangle$ and $|\Psi^{K'}_3\rangle$ order parameter (Figs. \ref{Fig3}(g) and (h)), which well match with the theoretical phase profile shown in Figs. \ref{Fig3}(c) and (d).
By selecting the K-point or K$'$-point reference, we can extract the order parameter profile of $|\Psi^{K}_3\rangle$ or $|\Psi^{K'}_3\rangle$ condensates respectively.
This vortex-antivortex lattice order is only seen above the threshold, and thus originating from the K- and K$'$-points condensates \cite{Sup4}.
\begin{figure}
\begin{center}
\includegraphics[scale=0.25]{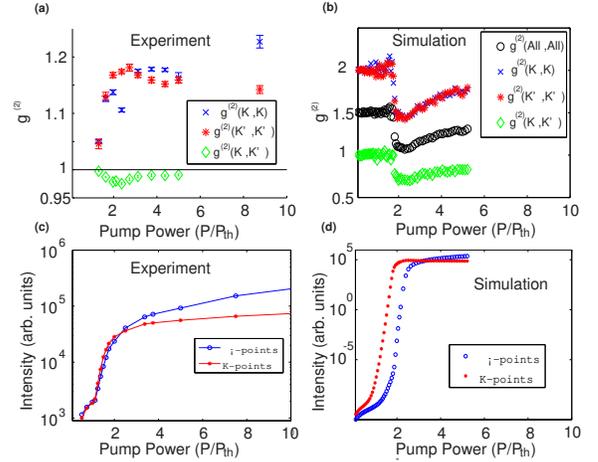}
\caption{
(a) Experimental results for the normalized second order auto correlation, $g^{(2)}({\rm K},{\rm K},\tau =0)$, $g^{(2)}({\rm K'},{\rm K'},\tau =0)$ and the cross correlation $g^{(2)}({\rm K},{\rm K'},\tau =0)$.
(b) Simulation results for  $g^{(2)}({\rm K},{\rm K},\tau =0)$, $g^{(2)}({\rm K'},{\rm K'},\tau =0)$ and $g^{(2)}({\rm K},{\rm K'},\tau =0)$ as well as $g^{(2)}({\rm All},{\rm All},\tau =0)$.
(c) (d) Experimental (c) and simulation (d) results for the pump power dependence of the intensity at $\Gamma$-point and the K-point condensate.
(c) The peak intensity values at the $\Gamma$-point and one K-point are plotted.
(d) The $\Gamma$-point intensity ($|\Psi _G|^2$) and the K-point total intensity ($|\Psi _K|^2 + |\Psi _{K'}|^2$) are plotted.
}
\label{Fig4}
\end{center}
\end{figure}

Finally we explore the dynamics of the mode competition between the degenerate $|\Psi^{K}_3\rangle$ and $|\Psi^{K'}_3\rangle$ condensates.
Figure \ref{Fig4}(a) shows the normalized second order auto and cross correlation function of $|\Psi^{K}_3\rangle$ and $|\Psi^{K'}_3\rangle$ condensates with the pump power dependence of the $\Gamma$-point and K(and K$'$)-points condensate intensity (Fig. \ref{Fig4}(c)).
$g^{(2)}({\rm K},{\rm K},\tau =0)$ ($g^{(2)}({\rm K'},{\rm K'},\tau =0)$) is the measured auto-correlation function of one K(K$'$)-point selected by the pinhole, while $g^{(2)}({\rm K},{\rm K'},\tau =0)$ is the measured cross correlation function between one K-point and one K$'$-point.
In both cases, the band pass filter with 1 nm band width is used to filter out the residual signals and Si avalanche photodiodes are used for photon-counting. 
We can see the anti-bunching behavior between $|\Psi^{K}_3\rangle$ and $|\Psi^{K'}_3\rangle$ condensates.
The degree of the anti-bunching has the maximum at P/P${\rm _{th}} \sim$ 2, where the K(and K$'$)-points condensates are dominant in the system.
With increasing the pump power, the $\Gamma$-point signal becomes dominant and the anti-bunching becomes smaller.
We relate this intensity anticorrelation to the mode competition resulting from the stochastic formation of $|\Psi^{K}_3\rangle$ and $|\Psi^{K'}_3\rangle$ condensates.
In each pulse, the $|\Psi^{K}_3\rangle$ and $|\Psi^{K'}_3\rangle$ condensates have the common particle reservoir.
The LP relaxation is first initiated by the spontaneous scattering through the LP-LP and LP-phonon scattering.
When the population at one state first reaches the quantum threshold, the stimulated scattering to that state is turned on and the relaxation into that state is more enhanced than the other, while the total particle number is determined by the pump power above the threshold.
Thus, the intensity anticorrelation between the $|\Psi^{K}_3\rangle$ and $|\Psi^{K'}_3\rangle$  condensates occurs.

We also simulate this stochastic condensates formation using the complex-number Langevin equations, similar to those for the polariton condensate order parameter with spins \cite{Read,Ohadi}.
We extend the model to include the $\Gamma$-point condensate.
We assume that the relaxation from the $|\Psi^{K}_3\rangle$ and $|\Psi^{K'}_3\rangle$ condensates to the ground state is mediated by the thermal reservoir because multi-phonon scattering processes are included.
The set of equations are 
\begin{eqnarray}
\frac{d}{dt}\Psi _{\sigma} &=& \frac{1}{2}\left[ -\Gamma _{\sigma}- \gamma _{c} \left( |\Psi_{\sigma}|^2 + 1 \right) + \gamma N_{R} \right] + \theta _{\sigma} (t), \\
\label{eq:one}
\frac{d}{dt}\Psi_{G} &=& \frac{1}{2} \left[ -\Gamma _{G} + \gamma _{G}N_{R} + \gamma _{c} \left( |\Psi_{K}|^2 + |\Psi_{K'}|^2 \right) \right] \nonumber\\
    &+& \theta _{GR} (t) + \theta _{GK} (t) + \theta _{GK'} (t), \\
\label{eq:two}
\frac{d}{dt}N_{R} &=& P - \Gamma _{R}N_{R} - \gamma N_{R} \left( |\Psi_{K}|^2 + 1 \right) \nonumber\\
    &-& \gamma N_{R} \left( |\Psi_{K'}|^2 + 1 \right) - \gamma _G N_{R} \left( |\Psi_{G}|^2 + 1 \right),
\label{eq:three}
\end{eqnarray}
where $\sigma = K$ or $K'$, $\Psi_{K}$, $\Psi_{K'}$ and $\Psi_{G}$ are the order parameter for $|\Psi^{K}_3\rangle$, $|\Psi^{K'}_3\rangle$ and the ground state (the $\Gamma$-points) condensates respectively \cite{Sup5}.
Because our pump laser is a pulsed one, we put $N_{R}(0)=\int dtP(t)$ and set P=0 for all time \cite{Read}. 
The dynamical threshold condition $N_{R}(0)=\int P_{\rm th} = \Gamma _K / \gamma$ for the case without the $\Gamma$-point condensate \cite{Rubo} is used here, though the presence of the $\Gamma$-point condensate may change the actual threshold.
We numerically solve the equations using a (fifth-order) Adams-Bashforth-Moulton predictor-corrector method and take 1000 samples for each power \cite{Note2}.
The simulation results are shown in Figs. \ref{Fig4}(b) and (d). 

In the simulation, at $P/P_{\rm th}<1.9$, $g^{(2)}(K,K)$ and $g^{(2)}(K',K')$ show the thermal state behavior, $g^{(2)}(\tau=0) \sim 2$ \cite{Note3}.
$g^{(2)}({\rm All},{\rm All}) \sim 1.1$ at $P/P_{\rm th} \sim 1.9$ indicates that the order parameters are well defined at the K- and K$'$-points and the states are coherent states.
The stochastic relaxation induces the anticorrelation between $n_K$ and $n_{K'}$, and it appears as the bunching in $g^{(2)}(K,K)$ and $g^{(2)}(K',K')$ and the anti-bunching in $g^{(2)}(K,K')$.
$g^{(2)}({\rm All},{\rm All}) \sim 1.1$ also shows that there is an excess noise induced by the relaxation process from the reservoir.
This excess noise makes the degree of the anti-bunching in $g^{(2)}(K,K')$ smaller than the degree of the bunching in $g^{(2)}(K,K)$ and $g^{(2)}(K',K')$.
We note that the thermal and quantum depletion \cite{Pitaevskii}, which are not included in this model, are other important sources for the excess noise in experiments \cite{Horikiri}.
With increasing the pump power, the $\Gamma$-point condensate become stronger than the K-point and K$'$-point condensates (Fig. \ref{Fig4}(d)), and the excess noise is also increasing (Fig. \ref{Fig4}(b)).
Here the relaxation from the $|\Psi^{K}_3\rangle$ and $|\Psi^{K'}_3\rangle$ to the $\Gamma$-point is comparable to the stimulated scattering from the reservoir to $|\Psi^{K}_3\rangle$ and $|\Psi^{K'}_3\rangle$.
Thus the noise effect becomes bigger and brings the intensity fluctuation to the $|\Psi^{K}_3\rangle$ and $|\Psi^{K'}_3\rangle$ condensates.
The small degree of the anti-bunching in experiments indicates that not all reservoir particles are common for $|\Psi^{K}_3\rangle$ and $|\Psi^{K'}_3\rangle$ .
Because of the momentum conservation in the relaxation process, some LPs depending on their momentum can only be scattered to either $|\Psi^{K}_3\rangle$ or $|\Psi^{K'}_3\rangle$.
This simulation qualitatively reproduces the experimental result.

We briefly note that although the LP-LP interaction cannot change the LP population in $|\Psi^{K}_3\rangle$ and $|\Psi^{K'}_3\rangle$ due to the wave-function parity symmetry, the phonon scattering or impurity scattering can in principle transfer the LPs from  $|\Psi^{K}_3\rangle$ to $|\Psi^{K'}_3\rangle$ and vice versa, and they may contribute the intensity anticorrelation.
We also note that the LP-LP repulsive interaction may induce the coherence between $|\Psi^{K}_3\rangle$ and $|\Psi^{K'}_3\rangle$ condensates.
Such dynamics between the $|\Psi^{K}_3\rangle$ and $|\Psi^{K'}_3\rangle$ condensates remains for the future study.

Our results show the peculiar characteristics of polariton condensates in two degenerate orbital states, where the stochastic relaxation process plays an important role.
We also show that it is possible to form the novel phase order, which is vortex-antivortex lattice order, in the polariton condensates originating from the single particle wavefunction in honeycomb lattice potentials. 

% acknowledgement.tex                            8 August 2012 
%
We would like to thank N. Kumada, C. Wu, M. Kuwata-Gonokami, Y. Shikano and E. Abe for
the fruitful discussion. We also thank M. Ueki for his technical expertise and full
support in the fabrication process.
We acknowledge JSPS through its FIRST program, Project for Developing Innovation Systems of the Ministry of  Education, Culture, Sports, Science and Technology (MEXT), Japan, Navy/SPAWAR Grant N66001-09-1-2024 and the State of Bavaria.
%
   % input acknowledgement

\end{document}